\documentclass[12pt]{iopart}

\usepackage{amssymb,graphicx,color,braket,hyphenat,makeidx,xcolor,dsfont,subfigure}
\usepackage{color}
\usepackage[ocgcolorlinks,colorlinks=true,linkcolor=blue,citecolor=red,linktocpage=true]{hyperref}

\newcommand{\blue}{\color{\blue}}

\newcommand{\ii}{i}  

\newcommand{\eqref}[1]{(\ref{#1})}
\newcommand{\text}[1]{\rm{#1}}
\newcommand{\comm}[2]{\left[#1,#2\right]}

\begin{document}

\title{Exploiting coherence for quantum thermodynamic advantage}

\author{Kenza Hammam$^{1,2,3}$, Heather Leitch$^1$,\\
 Yassine Hassouni$^2$, Gabriele De Chiara$^1$}




\address{$^1$ Centre for Quantum Materials and Technology, School of Mathematics and Physics,
Queen’s University Belfast, Belfast BT7 1NN, United Kingdom}
\address{$^2$ Equipe des Sciences de la Mati\`ere et du Rayonnement (ESMaR), Facult\'e des Sciences, Universit\'e Mohammed V, Av. Ibn Battouta, B.P. 1014, Agdal, Rabat, Morocco.}
\address{$^3$ International Center for Theoretical Physics ICTP, Strada Costiera 11, I-34151, Trieste, Italy}

\begin{abstract}
The introduction of the quantum analogue of a Carnot engine based on a bath comprising of particles with a small amount of coherence initiated an active line of research on the harnessing of different quantum resources for the enhancement of thermal machines beyond the standard reversible limit, with an emphasis on non-thermal baths containing quantum coherence. In our work, we investigate the impact of coherence on the thermodynamic tasks of a collision model which is composed of a system interacting, in the continuous time limit, with a series of coherent ancillas of two baths at different temperatures.  Our results show the advantages of utilising coherence as a resource in the operation of the machine, and allows it: (i) to exhibit unconventional behaviour such as the appearance of a hybrid refrigerator, capable of simultaneous refrigeration and generation of work, and (ii) to function as an engine or a refrigerator with efficiencies larger than the Carnot bound. Moreover, we find an effective upper bound to the efficiency of the thermal machine operating as an engine in the presence of a coherent reservoir. 
\end{abstract}

\maketitle

\section{Introduction}
Due to the nanotechnological progress and the increasing interest in quantum systems, heat engines are no longer limited to the size of steam engines from the industrial revolution \cite{Carnot1824}. Their miniaturisation is currently leading to the emergence of thermal machines that harness quantum effects to operate. The question of whether new thermodynamic characteristics may emerge from quantum features and aid thermodynamic tasks has long been debated. Answering this challenge might bring genuine quantum advantages in thermodynamics. This outlook fuelled a plethora of research papers on the use of quantum properties, such as quantum correlations \cite{ZhangPRA2007,DillenschneiderEPL2009,WangPRE2009,ParkPRL2013,AltintasPRE2014,Brunner2014,DeffnerPRE2015,BarriosPRA2017,HewgillPRA2018,DeChiaraPRR2020,KhandelwalNJP2020,TavakoliPRA2020,JosefssonPRB2020,Barrios2021,BresquePRL2021} and quantum coherence \cite{ScullyScience2003,ScullyPNAS2011,RahavPRA2012,BrandnerNJP2015,KilloranJCP2015,AltintasPRA2015,UzdinPRX2015,turkPRE2016,Uzdin2016,TurkEPL2017,Niedenzu2018,DorfmanPRE2018,PtaszynskiPRB2018,LatuneSR2019,CamatiPRA2019,ThingnaPRE2019,Um2021,LatuneQST2019}, in the design of quantum engines. Their genuine quantum effects have been observed in recent experiments \cite{KlatzowPRL2019,PetersonPRL2019}.

In recent years, quantum resource theories have been established to study and characterise the practicality of quantum coherence effects \cite{Streltsov2017}. In the context of thermodynamics, coherence was shown to be an essential ingredient for optimal charging of quantum batteries \cite{AlickiPRE2013,CampaioliPRL2017,FerraroPRL2018,AndolinaPRL2019,KaminPRE2020,SeahPRL2021,kim2021quantum,Wenniger2022},  it helps the transfer of energy in photosynthetic complexes \cite{MohseniJCP2008,CarusoPRA2010} and leads to the emergence of interesting phenomena like heat flow reversals without reversing the arrow of time \cite{LatunePRR2019}. Coherence has been investigated as a promising candidate for designing thermal machines in which the working medium is a quantum system in contact with reservoirs at different temperatures  and as a resource for exploring the fundamental limit of their efficiencies  \cite{UzdinPRA2016,DorfmanPRE2018,latuneEPJ2021}. 
As a matter of fact, it is of great importance to consider scenarios where one can model and manipulate non-equilibrium reservoirs. By employing them, one expects to overcome the standard thermodynamic limits as proposed in the case of quantum measurement induced \cite{ElouardPRL2017,talknerPRE2017}, squeezed \cite{KlaersPRX2017,Niedenzu2018,CorreaSR2014,HuangPRE2012,RossnagelPRL2014,ManzanoPRE2016} and coherent \cite{ScullyScience2003,dagEn2016,NiedenzuIOP2016,turkPRE2016,HaiPRE2014,YuEn2021,Opatrarxiv2021,GuffPRE2019,RomanIOP2019,Khanarxiv2021} engineered reservoirs. Analysing the machine’s operation in both cases of thermal and coherent reservoirs represents the most genuine way to explore novel aspects that may be induced by coherence. For instance, it was reported in Ref.~\cite{MaArxiv2021} that coherence injected in a finite sized reservoir interacting with a system can quantitatively modify the second law of thermodynamics and serves as a useful supply for enhancing the output efficiency of heat engines functioning between two coherent reservoirs.

For conventional system-reservoir models with a large continuum of modes, injecting coherence may become technically onerous \cite{LatuneQST2019}. A more versatile alternative that may lower these practical limitations is offered by quantum collision models (CM) \cite{ciccarelloarxiv2021,CampbellEPL2021,ManzanoPRX2022,RomanPRA2021,leitch2021,SeahPRE2019,GabrielePRL2021,DeChiaraPRR2020,Barra2015,Strasberg2016}. Because of the simplicity of their mechanism, they are well-qualified to address the thermodynamics of coherent engineered reservoirs \cite{LiPRE2018,ManatulyPRE2019,RodriguesPRL2019}. The basic formulation of CMs consists of depicting the environment as a large ensemble of elementary components or sub-units, often called \textit{ancillas}, that interact sequentially with a system $S$. The system-ancilla joint dynamics are described by collisions that may or may not preserve energy. The conditions for a memoryless Markovian behaviour of the open quantum system evolution imply that the ancillas are initially prepared in a product state (uncorrelated) and that every ancilla collides only once with $S$. In addition to being an interesting asset to study non-thermal environments, this microscopic formalism is highly suited to model specific experiments based on micromasers \cite{FiliPRA1986,HarocheOxford2006}, nuclear magnetic resonance NMR \cite{NadjaSR2016}, superconducting quantum computers \cite{GarciaPerez2020,CattaneoPRL2021,MeloPRA2022,Cattaneo2022}  and all-optical settings \cite{JinPRA2015,CuevasSR2019,BernadesSR2015}.

Although there is an active line of ongoing research on quantum CMs, the different effects of exploiting coherence in the quest of enhanced quantum thermal machines are little explored so far. On the other hand, it was a motivation for a recent derivation of the thermodynamic behaviour of an autonomous quantum thermal machine \cite{HammamIOP2021} that operates via collisions with a series of qubits which are initially prepared with energetic coherence, as well as a basic setup of a system interacting with coherent environmental ancillas \cite{RodriguesPRL2019}. 

In this work, we study a thermal machine consisting of a single qubit as a working substance and two reservoirs modelled by CMs. We carefully analyse the thermodynamics of the system's evolution in the presence of an infinitesimal amount of coherence in the reservoirs and consider the non-equilibrium steady state of the system \cite{GuarnieriPRL2018,GuarnieriPLA2020}.  We split the heat flows into a coherent contribution, arising because of the quantum coherence in the baths,
and an incoherent contribution corresponding to the typical dissipation term.  As a consequence of having non-resonant collisions, power can instead be divided into a term originating from coherence and a term that stems from the collisions.  
Furthermore, consuming coherences from the reservoirs allows the machine to perform thermodynamic tasks in classically forbidden regimes with efficiencies beyond the Carnot bound of the corresponding thermal reservoirs without coherence. We find the effective bound on the efficiency of the machine operating as an engine in the presence of quantum coherence. For a cold coherent reservoir, the device is capable of simultaneous refrigeration and generation of work, providing a kind of device dubbed  \textit{hybrid refrigerator} \cite{ManPRR2020,NiedenzuIOP2016}. 

The paper is structured as follows: After introducing the coherent CM setup and its dynamics in Sec. \ref{sec:model}, we lay out the different results in the case of harnessing coherence in the cold and the hot bath in Sec. \ref{sec:results} as well as its influence on the operational regimes of the machine and its efficiency at maximum power. Finally, we provide some concluding remarks in Sec. \ref{sec:conclusion}.

\section{Model}
\label{sec:model}

\subsection{Evolution}
\begin{figure}[t]
\begin{center}
\includegraphics[width=0.7\columnwidth]{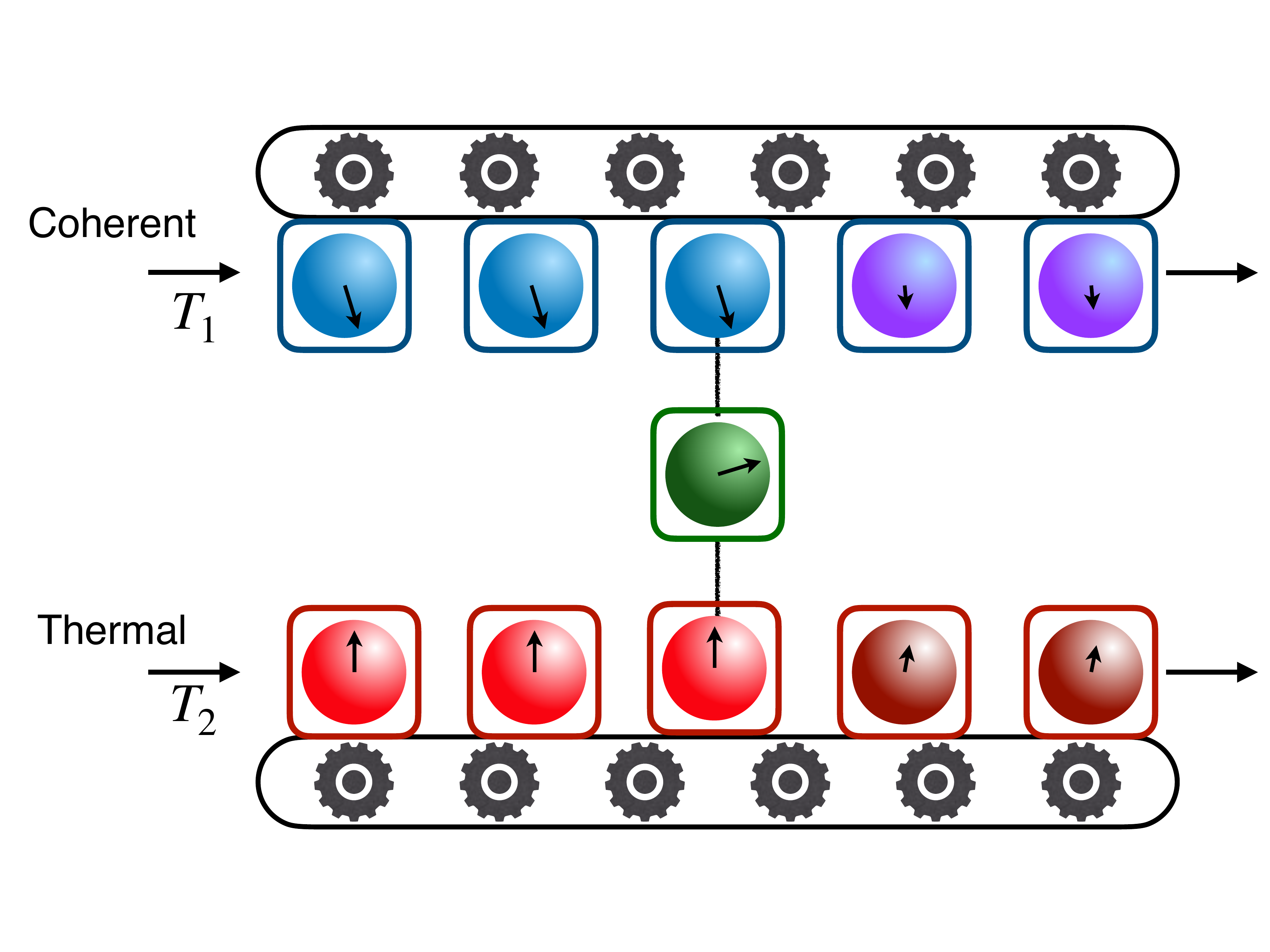}
\caption{Sketch of our setup: a qubit (centre, green) is coupled to two environments modelled with the collision model where the arrow represents its Bloch vector. The ancillas in one of the two environments (bottom, red) are initially in a thermal state as represented by a vertical Bloch vector. The ancillas of the other environment have initially some coherence in the energy eigenbasis as represented by a tilted initial Bloch vector. After the collisions, the ancillas are modified as represented by a different colour and Bloch vector.}
\label{fig:setup}
\end{center}
\end{figure}

We consider a system described by a Hamiltonian $H_S$ and interacting with an environment modelled with the collision model. The environment consists of $N_B$ baths each containing an infinite ensemble of identically prepared ancillas which interact sequentially with the system for a time $\tau$ and then are discarded, see Fig.~\ref{fig:setup}. We assume that there are no initial system-environment correlations.
The reduced dynamics of the system after one collision can be written as ($\hbar=1$):
\begin{equation}
\rho'_S = \tr_E \left\{e^{-i \tau H_{\rm tot}} \rho_S \otimes \rho_E e^{i \tau H_{\rm tot}} \right\}
\end{equation}
where $\rho_S$ $[\rho'_S]$is the state of the system before [after] the collision, 
\begin{equation}
\rho_E =\bigotimes_{i=1}^{N_B} \rho_{E,i}
\end{equation}
is the state of the environment and we do not assume any initial correlations between the baths, and
\begin{equation}
H_{\rm tot}=H_S+H_E+H_{SE},
\end{equation}
is the total system-environment Hamiltonian.
The total Hamiltonian of the environment is the sum of the corresponding Hamiltonian $H_{E,i}$ for each bath:
\begin{equation}
H_{E} = \sum_{i=1}^{N_B} H_{E,i}.
\end{equation}
 We assume the system-environment interaction to be of the general form $H_{SE}=\sum_{i=1}^{N_B} H_{SE_i}$ and
 \begin{equation}
 \label{eq:HSE}
H_{SE_i}=
\sum_k \frac{g_{i,k}}{\sqrt \tau} \left (S_{i,k}^\dagger A_{i,k} + S_{i,k} A_{i,k}^\dagger \right),
\end{equation}
where $g_{i,k}$ are the coupling constants between the system and the ancilla from bath $i$. The index $k$ lists the different system's operators interacting with a corresponding ancilla. The operators $S_{i,k}$ and $A_{i,k}$ pertain to the system and ancilla's Hilbert space, respectively. Moreover, we also assume the operators $A_{i,k}$ to be eigenoperators of the ancilla's Hamiltonian, such that:
\begin{equation}
H_{E,i} = \sum_k \omega_{i,k} A^\dagger_{i,k} A_{i,k}.
\end{equation}
As we will see, the factor $\sqrt \tau$ in Eq.~\eqref{eq:HSE}, though not necessary, ensures consistency when taking the continuous limit $\tau\to 0$.

We now assume that the ancillas are prepared in a thermal state at temperature $T_i$  with a small coherence term ($\beta_i=T_i^{-1}, k_B=1$):
\begin{equation}
\rho_{E,i} = \rho^{\rm th}_{E,i} +\sqrt \tau \epsilon_i \chi_{E,i}
\end{equation}
where $ \rho^{\rm th}_{E,i} = {e^{-\beta_i H_{E,i}} }/{\tr \left[e^{-\beta_i H_{E,i}} \right]}$ and $\epsilon_i$ quantifies the ancilla's quantum coherence.

In the continuous limit, $\tau\to 0$, following Ref.~\cite{RodriguesPRL2019,DeChiaraNJP2018}, the evolution of the system's reduced density matrix is ruled by the following Markovian master equation:
\begin{equation}
\dot\rho_S = -i[H_S+G_S,\rho_S]+\sum_{i=1}^{N_B} D_i(\rho_S)
\end{equation}
where the effective Hamiltonian correction is given by
\begin{eqnarray}
\label{eq:GS}
G_S&=& \tr_{E}\left(H_{SE}\rho_{E}\right )
\\
&=&\sum_{i=1}^{N_B}  \epsilon_i \sum_k g_{i,k}\left\{S_{i,k}^\dagger \tr_{E} \left[ \chi_{E,i} A_{i,k}\right] + \rm{h.c.}\right\}.
\end{eqnarray}
The dissipators are defined by:
\begin{equation}
D_i(\rho_S) = \sum_k \gamma^{-}_{i,k} \mathcal L[S_{i,k},\rho_S]
+\gamma^{+}_{i,k} \mathcal L[S^\dagger_{i,k},\rho_S]
\end{equation}
where the Lindbladian is defined as:
\begin{equation}
\mathcal L [S,\rho]  = 2 S\rho S^\dagger-\{S^\dagger S, \rho\}.
\end{equation}
The dissipation rates $\gamma^{-}_{i,k}=g^2_{i,k} \langle A_{i,k}A^\dagger_{i,k}\rangle$ and $\gamma^{+}_{i,k}=g^2_{i,k} \langle A^\dagger_{i,k}A_{i,k}\rangle$, where the averages $\langle \cdot\rangle$ are taken over the thermal part of the environmental state $\rho_{E,i}^{\rm th}$, fulfil  the local detailed balance condition $\gamma^{+}_{i,k}/\gamma^{-}_{i,k} = \exp(-\beta_i\omega_{i,k})$.

\subsection{Thermodynamic quantities}
Thermodynamic quantities can be calculated following Ref.~\cite{DeChiaraNJP2018,RodriguesPRL2019,DeChiaraPRR2020}. We define work in the usual way as arising from the time-dependence of the total Hamiltonian, see below. Other contributions to the change of internal energy come from energy exchanges between the system and the environment and  will be collectively categorised as heat.

Therefore, the heat current flowing from bath $i$ is obtained from the energy change of the corresponding ancilla before and after the collision:
\begin{equation}
\label{eq:Qdot}
\dot Q_i =-\lim_{\tau\to 0} \frac 1\tau \tr\left[H_{E,i} \Delta \rho\right],
\end{equation}
where we have defined the change in the total density matrix of the system plus environment:
\begin{equation}
\Delta\rho=e^{-i \tau H_{\rm tot}} \rho_S \otimes \rho_E e^{i \tau H_{\rm tot}} -\rho_S \otimes \rho_E.
\end{equation}
Expanding Eq.~\eqref{eq:Qdot}, and following the detailed derivation presented in \ref{app:derivations}, we may split the heat flow from bath $i$ into coherent and  incoherent contributions:
\begin{equation}
\dot{Q}_i = \dot{Q}_{i,\text{coh}}+\dot{Q}_{i,\text{inc}}.
\end{equation}
The coherent contribution of the heat arises because of the initial coherence in the ancilla and reads:
\begin{equation}
\label{eq:Qdotcoherent}
\dot{Q}_{i,\text{coh}} = -\lim_{\tau\to 0} i \tr \left([G_{E,i},H_{E,i}] \rho_{E,i}\right ),
\end{equation}
where, in analogy to the operator $G_S$, we have defined
\begin{eqnarray}
\label{eq:GEi}
G_{E,i} &=& \tr_{\overline{E_i}} \left(H_{SE_i}\rho_S \right )
\end{eqnarray}
where $\tr_{\overline{E_i}}$ is the partial trace on the system and all the environments except the $i$th.
The heat current's incoherent contribution depends on the typical double-commutator structure:
\begin{eqnarray}
\label{eq:Qdotincoherent}
\dot{Q}_{i,\text{inc}} &=& 
\frac{1}{2} \lim_{\tau\to 0}\tr\left(\tau [H_{SE},[H_{SE},H_{E,i}]] \rho_S \otimes \rho_E\right).
\end{eqnarray}
Notice, however, that the last expression does not necessarily reduce to the common expression involving the system's expectation value of the dissipator applied to the system's Hamiltonian $\tr [D_i(H_S) \rho_S]$ unless the ancillas are resonant with the system's transition they are coupled to. In the next section, we will consider a situation in which this resonant condition is not met.

The total power that needs to be injected or extracted from the system is
\begin{equation}
\label{eq:Wdot}
\dot W= \lim_{\tau\to 0} \frac 1\tau \tr\left[\left(H_{S}+H_E\right) \Delta \rho \right].
\end{equation}
Notice that if we define the rate of change of the internal energy:
\begin{equation}
\dot U_S= \lim_{\tau\to 0} \frac 1\tau \tr\left[H_{S} \Delta \rho \right],
\end{equation}
then the first law of thermodynamics is automatically satisfied:
\begin{equation}
\dot U_S = \dot W +\sum_{i=1}^{N_B} \dot Q_i.
\end{equation}
The total power can be split into a coherent and a collisional contribution:
\begin{equation}
\dot{W} =    \dot{W}_{\text{coh}}+ \dot{W}_{\text{coll}}.
\end{equation}
The coherent contribution for the power arises only when the ancilla has some initial coherence and its expression reads:
\begin{equation}
 \dot{W}_{\text{coh}}=\ii   \lim_{\tau\to 0}\tr\left( [H_{SE},H_S+H_E] \rho_S \otimes \rho_E\right).
\end{equation}
The power due to the collision
\begin{equation}
\dot{W}_{\text{col}} = - \frac{1}{2} \lim_{\tau\to 0}\tr\left(\tau [H_{SE},[H_{SE},H_S+H_E]] \rho_S \otimes \rho_E\right),
\end{equation}
arises for local (but not global) master equations when the system-environment coupling does not satisfy local energy conservation, in other words, when $[H_{SE},H_S+H_E]\neq 0$ \cite{HewgillPRR2021}.

The splitting of heat and work considered here is in line with other studies on quantum thermodynamics with non-equilibrium reservoirs, see for instance Ref.~\cite{RodriguesPRL2019}. However, we remark that this splitting is different, although physically equivalent, to the one considered in Ref.~\cite{RodriguesPRL2019} in which the condition $[H_{SE},H_S+H_E]= 0$ was assumed. As a consequence, the term that we dub $\dot Q_{i, \text {coh}}$ was interpreted there as a power term arising from coherence while the power terms we dub $ \dot{W}_{\text{coh}}$ and $ \dot{W}_{\text{coll}}$ were identically zero. 

In this paper, all terms arising from the energy change of the environment have been identified as heat terms, coherent and incoherent. The remaining terms in the energy balance arise from the time-dependence of the total Hamiltonian have been identified as power contributions which include a term $ \dot{W}_{\text{coh}}$ only arising because of the ancilla's coherence and a term $\dot{W}_{\text{col}}$ only arising because of the ``locality" of the master equation as discussed in \cite{DeChiaraNJP2018}.

If the system's Hamiltonian is explicitly time-dependent there will also be an associated power term proportional to $\partial H_S/\partial t$. We will not consider this additional term in this work.
We are assuming the sign convention such that power or heat currents are positive when energy flows into the system.
Depending on the signs of these quantities the device exhibits different functionings as discussed in the next section. 

\subsection{Entropic quantities}
We now pass to discuss the change in entropic quantities and the second law of thermodynamics.
During a collision, the ancillas may lose or gain quantum coherence related to the off-diagonal entries of their density matrices in their energy eigenbasis. By using the definition of the relative entropy of coherence \cite{Streltsov2017,BaumPRL2014}
\begin{equation}
\mathcal{C}(\rho) = S(\rho^{\text{d}}) - S(\rho),
\end{equation}
where $S(\rho)=-\tr\rho\log\rho$ is the von Neumann entropy and $\rho^{\text{d}}$ is the diagonal part of $\rho$ in a given basis, we can measure the rate of change in relative entropy of coherence of the ancillas before and after the collision:
\begin{equation}
 \dot\mathcal{ C} (\rho) = \lim_{\tau\to 0}\frac{\mathcal{C}(\rho') - \mathcal{C}(\rho)}{\tau}.
\end{equation}
Using perturbation theory and taking the limit of $\tau \to 0$, we find the following relation between the relative entropy, coherent heat flow and change in relative entropy of coherence of an ancilla
\begin{equation}
\label{eq:relative_entropy}
\dot{S}(\rho_{E,i}' || \rho_{E,i}) = \beta_i \dot{Q}_{i,\text{coh}} + \dot{\mathcal{C}}(\rho_{E,i}) ,  
\end{equation}
where we have defined the relative entropy as $S(\rho' || \rho) = \text{Tr}[\rho' \log \rho']-\text{Tr}[\rho' \log \rho]$.
This relation extends the result found in Ref.~\cite{RodriguesPRL2019} to a more general scenario: there the relative entropy was related to the coherent power and not to the coherent heat.
Since the relative entropy in Eq.~\eqref{eq:relative_entropy} cannot be negative, we obtain a  lower bound for the coherent heat flow from each bath, which can be interpreted as a modified local second law of thermodynamics
\begin{equation}
\label{eq:bound1}
\beta_i \dot{Q}_{i,\text{coh}} \geq - \dot{\mathcal{C}}(\rho_{E,i}).   
\end{equation}

While the previous expression concerns each bath individually, we can also obtain a modified second law for the whole environment. Because the initial state of the ancilla's, $\rho_E$, is a product state, we may write the total relative entropy of the bath as 
\begin{equation}
S(\rho_E'||\rho_E) = \mathcal{I}(\rho_E') + \sum_{i}  S(\rho_{E,i}'||\rho_{E,i})  
\end{equation}
where we have defined the state of the environment after each collision
\begin{equation}
\rho_E' = \text{Tr}_S [\rho']
\end{equation}
and the total mutual information as
\begin{equation}
\mathcal{I}(\rho_E')  = \sum_{i=1}^{N_B} S(\rho_{E,i}') - S(\rho_{E}'),
\end{equation}
while $\mathcal{I}(\rho_E)=0$. 
We thus find:
\begin{eqnarray}
\dot S(\rho_E' || \rho_E) &=&  \dot\mathcal{ I}(\rho_E')+ \sum_{i=1}^{N_B} \dot S(\rho_{E,i}'||\rho_{E,i}) =
\\
&=&\dot\mathcal{ I}(\rho_E')+ \sum_{i=1}^{N_B}\beta_i \dot{Q}_{i,\text{coh}} + \dot{\mathcal{C}}(\rho_{E,i})\geq 0,
\end{eqnarray}
which represents a global modified second law.

Classes of modified versions of the second law of thermodynamics stemming from coherence in a system interacting with a large environment have been investigated in the literature, see e.g. \cite{RichensPRE2018,LostaglioNC2015,LobejkoNC2021} whereas, for coherent reservoirs, they remain not widely explored \cite{RodriguesPRL2019,Strasberg2016}. Our modified second law demonstrates that the coherent heat contribution is constrained by the loss of coherence in the auxiliary's state and that clearly shows that coherence is a resource to be harnessed to perform thermodynamic tasks in non-equilibrium processes. Ref.~\cite{MaArxiv2021}, obtained similar results but for a generic finite size reservoir with coherence.

The results of this section are quite general and only depend on a few assumptions, chiefly no initial system-environment correlations  or within the environment. In the next section we showcase them for the specific case of a one-qubit system coupled to two environments at different temperatures. 


\section{Results}
\label{sec:results}
We now specialise our problem to that of a single qubit in contact with two baths ($N_B=2$). We assume the qubit's Hamiltonian to be:
\begin{equation}
H_S = B \sigma_S^z.
\end{equation}
We also assume the ancillas to be qubits such that their Hamiltonians read:
\begin{eqnarray}
H_{E,i} =B_i \sigma^z_{E,i}, \;\; i=1,2.
\end{eqnarray}
Notice that the ancillas are {\it not} resonant with the qubit. This causes extra terms to appear in the power and heat expressions proportional to the non vanishing gap \cite{PiccionePRA2021}. 
The system-environment interaction Hamiltonian is assumed of the rotating-wave type with $S_i = \sigma_S^-$ and $A_{i}=\sigma^-_{E,i}$ (where we have therefore dropped the index $k$ as there is only one value).
For the state of the ancilla before the collision we choose:
\begin{equation}
\chi_{E,i} =\cos\phi_i \sigma^x_{E,i} + \sin\phi_i \sigma^y_{E,i}
\end{equation}
where the angle $\phi_i$ can be interpreted as the azimuth of the ancilla's Bloch vector. 

Under these assumptions, the master equation for the system's qubit becomes:
\begin{eqnarray}
\dot \rho_S &=& -i[H_S+G_S,\rho_S]+\gamma(n_1+n_2+2)\mathcal L [\sigma^-_S,\rho_S]
\nonumber
\\
&+&\gamma(n_1+n_2)\mathcal L [\sigma^+_S,\rho_S]
\label{eq:mequbit}
\end{eqnarray}
where the effective Hamiltonian correction is:
\begin{equation}
G_S = \sqrt{2\gamma} \sum_{i=1,2 } \epsilon_i \sqrt{2n_i+1}\left(\cos\phi_i \sigma^x_{S} + \sin\phi_i \sigma^y_{S}\right )
\end{equation}
and we have assumed equal rates $\gamma$ for the two baths such that: $g^2_{E,i}=2\gamma(2n_i+1)$, where $n_i=[\exp(2\beta_i B)-1]^{-1}$ is the thermal occupation in each ancilla. Notice that in the absence of environmental coherence ($\epsilon_i=0$), the master equation would simply correspond to that of a qubit in contact with a single effective bath with an average thermal occupation $n=(n_1+n_2)/2$ and in the long-time limit will equilibrate with this average thermal occupation.

The steady state of the system's qubit can be found by solving the equation $\dot \rho_S = 0$ and its analytical expression is reported in 
\ref{sec:appendix}. This shows that the steady state density matrix elements are affected by the presence of coherence, resulting in terms proportional to $\epsilon_1$ and $\epsilon_2$.

We now pass to the thermodynamic quantities, heat currents and power. To this end, to simplify the results we assume coherence only in one bath and set $\epsilon_2=0$. The case with coherence in both baths does not lead to additional qualitatively different scenarios considered here as proven in 
\ref{sec:appendixC}.

Detailed expressions for the coherent, incoherent and collisions contributions to heat and work for any state of the system can be found in \ref{sec:appendixB}.
Substituting the expressions for the steady state, found in 
~\ref{sec:appendix}, in Eqs.~\eqref{eq:Qdot}-\eqref{eq:Wdot}, we obtain for the heat currents and power (totalling their coherent, incoherent and collisional contributions):
\begin{eqnarray}
\dot Q_1 = B_1 V(\epsilon_1)
\label{eq:Q11}
\\
\dot Q_2 = -B_2 V(\epsilon_1)
\label{eq:Q21}
\\
\dot W = (B_2-B_1) V(\epsilon_1)
\label{eq:W1}
\end{eqnarray}
where the common factor is given by:
\begin{equation}
\label{eq:V1}
\fl V(\epsilon_1)=2\gamma\frac{B^2(n_1-n_2)+\gamma(1+n_1+n_2) 
\left[(n_1-n_2)(1+n_1+n_2)\gamma+(2n_1+1)\epsilon_1^2 \right] }
{(1+n_1+n_2)\left[B^2+(1+n_1+n_2)^2\gamma^2+(2n_1+1)\gamma\epsilon_1^2  \right]}.
\end{equation}
The fact that the thermodynamic quantities contain a common factor is a consequence of the system-environment exchange interaction Hamiltonian that we have assumed and which was found in related models, e.g. in Refs.~\cite{DeChiaraNJP2018,DeChiaraPRR2020,PiccionePRA2021}. 
An immediate consequence is that the ratios of these quantities, linked to the efficiency and coefficient of performance (COP), only depend on the ratios of the ancillas magnetic fields yielding Otto-like expressions. For instance, we obtain the efficiency
\begin{equation}
\eta=-\frac{\dot W}{\dot Q_{\rm in}} = 1-\frac{\min\{B_1,B_2\}}{\max\{B_1,B_2\}},
\end{equation}
where the input heat $\dot Q_{\rm in}$ is the sum of all positive heat contributions. 

Notice that the factor $V(\epsilon_1)$ contains explicitly the strength of coherence $\epsilon_1$ but not its phase. In the case of two baths with coherence, only their relative phase would appear in the expression, see 
\ref{sec:appendixC}. 
The factor $V(\epsilon_1)$ also contains the system's magnetic field $B$ and would appear also in the absence of bath coherence \cite{PiccionePRA2021}. In this case, this field plays the role of an effective detuning which increases the magnitude of the heat currents and  power but does not alter their signs and therefore the type of operating machine.



In the next two subsections we illustrate our results for the case in which the cold environment produces coherent ancillas ($T_1<T_2$) and the opposite case in which the hot environment is coherent ($T_2<T_1$).

\subsection{Coherence in the cold bath}
We start by assuming that the coldest bath has some initial coherence; therefore we assume that $T_1<T_2$. In this case, depending on the choice of the coherence strength $\epsilon_1$ and the magnetic fields $B_1$ and $B_2$, the setup can operate as different types of thermal machines, as summarised in Table~\ref{tab:functionings}.
\begin{table}[h]
\caption{Functionings for $T_1<T_2$.}
\begin{center}
\begin{tabular}{|c|c|}
\hline
$\dot W<0, \dot Q_1<0,\dot Q_2>0$ & engine
\\
\hline
$\dot W<0, \dot Q_1>0,\dot Q_2<0$ & hybrid refrigerator
\\
\hline
$\dot W>0, \dot Q_1<0,\dot Q_2>0$ & accelerator
\\
\hline
$\dot W>0, \dot Q_1>0,\dot Q_2<0$ & refrigerator
\\
\hline
\end{tabular}
\end{center}
\label{tab:functionings}
\end{table}

\begin{figure*}[t]
\begin{center}
\includegraphics[height=3.7cm]{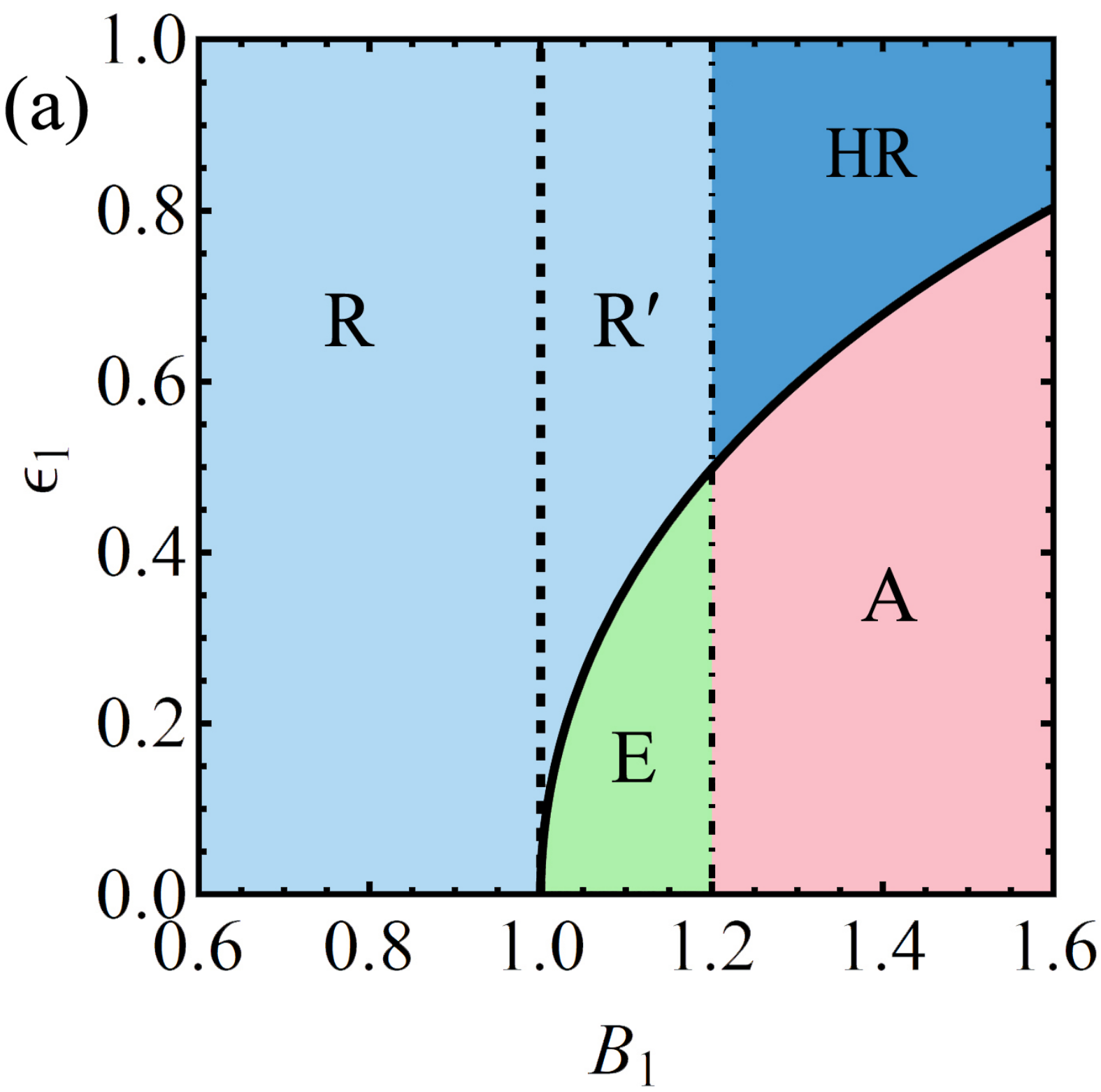}
\includegraphics[height=3.6cm]{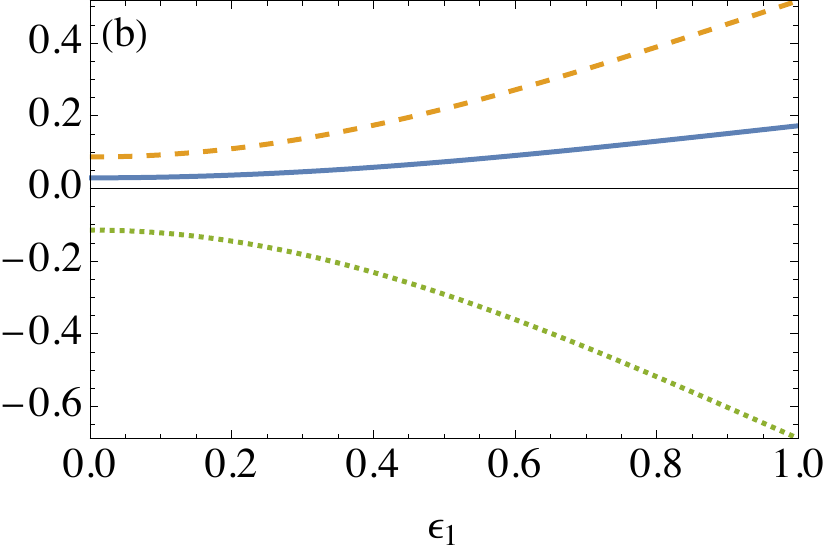}
\\
\includegraphics[height=3.6cm]{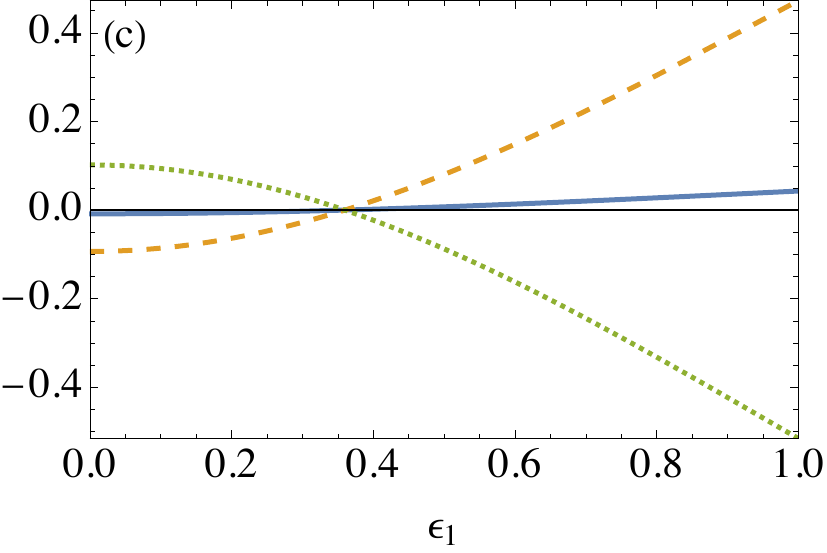}
\includegraphics[height=3.6cm]{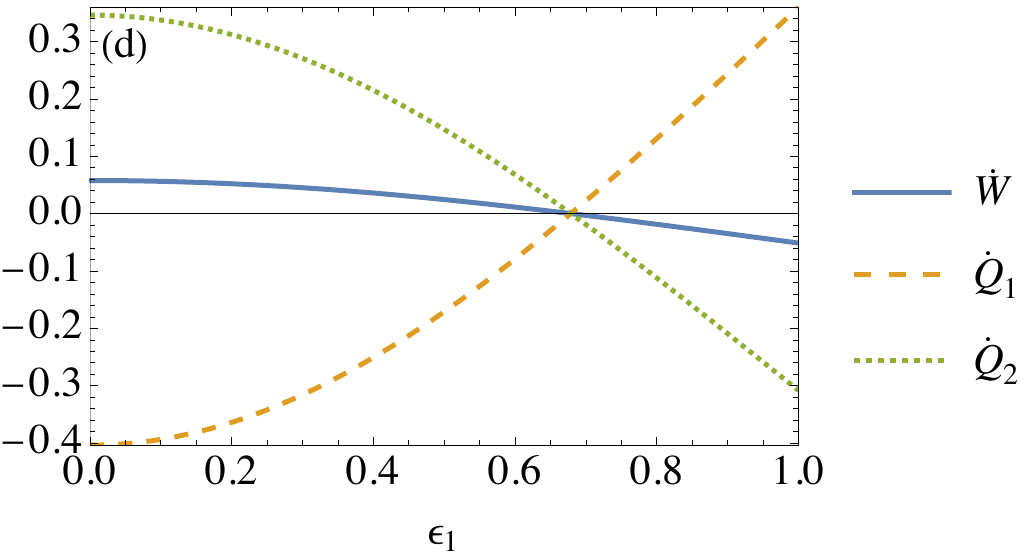}
\caption{Coherence in the cold bath: (a) Operational diagram for the different functionings of the thermal machine. Colour coding is as follows: R$\equiv$refrigerator, HR$\equiv$hybrid refrigerator, A$\equiv$accelerator, E$\equiv$engine. In the area denoted as R', the device works as a refrigerator with a COP larger than the Carnot value.  The vertical dashed line indicates the condition $n_1=n_2$; the vertical dotted-dashed line corresponds to the condition $B_1=B_2$; the curved solid line corresponds to the condition in Eq.~\eqref{eq:condition2}. Heat currents and power for (b) $B_1=0.9$, (c) $B_1=1.1$, (d) $B_1=1.4$. Other parameters: $B=1, B_2=1.2, T_1=2.5,T_2=3, \gamma=1$.}
\label{fig:coldbathcoherence}
\end{center}
\end{figure*}

Looking at Eqs.~\eqref{eq:Q11}-\eqref{eq:V1} we see that $\dot W$, but not  $\dot Q_1$ and  $\dot Q_2$, changes sign when 
\begin{equation}
\label{eq:condition1}
B_1=B_2,
\end{equation}
while all quantities, $\dot W, \dot Q_1$ and $\dot Q_2$, become zero when $V(\epsilon_1)$ cancels which occur, in the presence of coherence, when
\begin{equation}
\label{eq:condition2}
\epsilon^*_1=\frac{\sqrt{n_2-n_1}\sqrt{B^2+(1+n_1+n_2)^2\gamma^2}}{\sqrt{(1+2n_1)(1+n_1+n_2)\gamma}}.
\end{equation}
The two conditions \eqref{eq:condition1}-\eqref{eq:condition2} determine the functioning diagram reported in Fig.~\ref{fig:coldbathcoherence}(a). There, we see that all four regimes illustrated in Table~\ref{tab:functionings} appear for certain values of the parameters. In the panels (b,c,d) of Fig.~\ref{fig:coldbathcoherence} we plot the thermodynamic quantities $\dot W, \dot Q_1$ and $\dot Q_2$ along three cuts of the functioning diagram. The crossing points that appear in panels (c) and (d) of Fig.~\ref{fig:coldbathcoherence} correspond to effective Carnot points determined by Eq.~\eqref{eq:condition2} where all thermodynamic quantities go to zero and change sign.

\begin{figure}[htbp]
\begin{center}
\includegraphics[height=5cm]{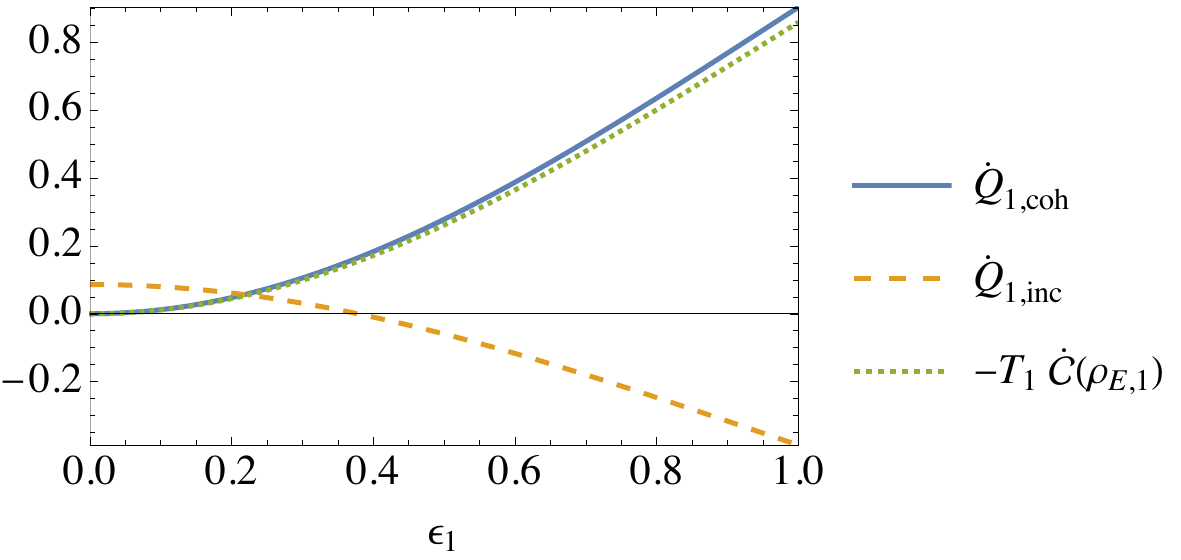}
\caption{Plot of the coherent contribution $\dot{Q}_{1,\text{coh}}$ (solid line) and incoherent contribution $\dot{Q}_{i,\text{inc}}$ (dashed line) heat flow from bath 1 against $\epsilon_1$. The change in relative entropy of coherence $ \dot{\mathcal{C}}(\rho_{E,1})$ for bath 1, multiplied by $-T_1$ (dotted line) is also plotted against $\epsilon_1$ to verify that the bound in Eq. \eqref{eq:bound1} holds. Parameters are as in Fig.~\ref{fig:coldbathcoherence} (b).}
\label{fig:Qcoherent}
\end{center}
\end{figure}

Figure \ref{fig:Qcoherent} shows how both the coherent and incoherent heat flows from bath $1$ vary with $\epsilon_1$. We see that at $\epsilon_1 \approx 0.2$, the contribution to the overall heat flow (from bath 1) due to coherence surpasses the incoherent heat flow.
The rate of change of coherence $ \dot{ \mathcal{C}}(\rho_{E,1})$ is also plotted in Fig.~\ref{fig:Qcoherent} providing evidence that the bound in Eq.~\eqref{eq:bound1} holds.

We now pass to discuss the efficiency and coefficient of performance of the thermal devices. As discussed in the previous section, these quantities are given by the Otto values:
\begin{equation}
\eta=1-\frac{B_1}{B_2},
\end{equation}
when operating as an engine and 
\begin{equation}
COP=\frac{B_1}{B_2-B_1},
\end{equation}
when operating as a refrigerator. 
In absence of coherence and for thermal baths, these quantities are smaller or equal than the corresponding Carnot values:
\begin{eqnarray}
\eta_C=1-\frac{T_1}{T_2}, \label{eq:carnot}
\label{eq:etac}
\\
COP_C=\frac{T_1}{T_2-T_1},
\end{eqnarray}
 with equality obtained only when the heat currents and power drop to zero. This is because, in absence of coherence ($\epsilon_1=0$), the condition $COP>COP_C$ is equivalent to the condition $n_1<n_2$ that corresponds to the functioning of the device as an engine. 
 However, in the presence of coherence, we see in Fig.~\ref{fig:coldbathcoherence} that the refrigerator regime survives in a classically ``forbidden" area for which $n_1<n_2$. 
 \begin{figure}[t]
\begin{center}
\includegraphics[height=5cm]{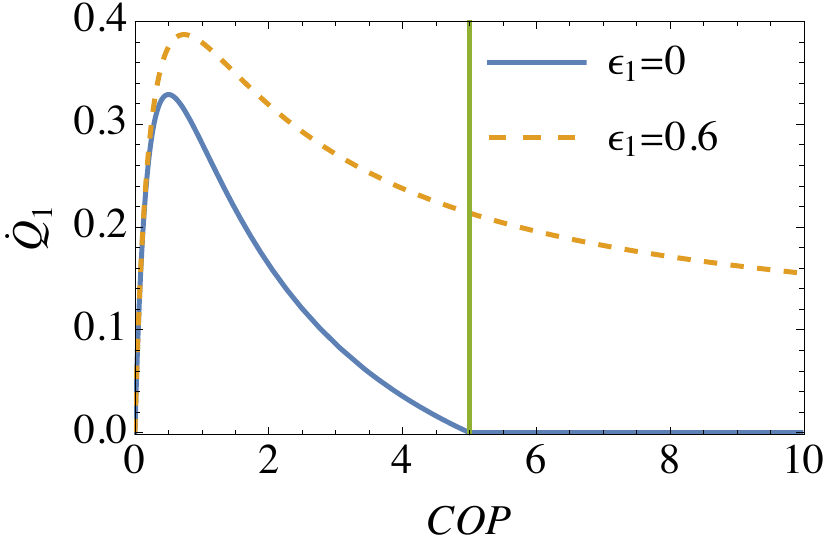}
\caption{Cooling power $\dot Q_1$ against the $COP$ for the device operating as a refrigerator. We compare the cases with no coherence (solid, $\epsilon_1=0$) and with coherence (dashed, $\epsilon_1=0.6$). Parameters as in Fig.~\ref{fig:coldbathcoherence} with $0<B_1<B_2$.}
\label{fig:COP1}
\end{center}
\end{figure}
In this area, labelled R$'$ in the diagram, the $COP$ is indeed larger   than the Carnot value $COP_C$ as evidenced in Fig.~\ref{fig:COP1}, where we plot the cooling power $\dot Q_1$ against the $COP$. We find that in the region R$'$ where $COP>COP_C$, the cooling power is nonzero, in contrast to the Carnot point where power is strictly null. Moreover, Fig.~\ref{fig:COP1} shows that the $COP$ at maximum cooling power is larger in the presence of coherence. The  $COP$ is not bounded and diverges at the transition between the functioning as a refrigerator and as a hybrid refrigerator for $B_1\to B_2$.
 
 Regarding the efficiency of the system as an engine, applying the same considerations as before, we find that the condition $\eta>\eta_C$ would correspond to  $n_2<n_1$. In Fig.~\ref{fig:coldbathcoherence}(a) however, we observe that the system never operates as an engine for $n_2<n_1$, even in the presence of coherence.
 
 Two more functionings appear in Fig.~\ref{fig:coldbathcoherence}(a): the accelerator and the hybrid refrigerator. For the accelerator, heat flows in the spontaneous direction (hot to cold) but this process is accelerated by positive work injected into the system and transformed into heat that is dissipated in the cold environment. The hybrid refrigerator instead does the opposite: it extracts heat from the coldest bath, converts part of it into work which can be extracted and dumps the rest into the hot bath. This apparently paradoxical functioning is made possible necessarily by the presence of quantum coherence in the cold bath which acts as an extra source of work.
 
 Summarising, when coherence is present in the cold bath, there is a region of parameters for which the system operating as a refrigerator has a larger, in principle unbounded, $COP$ than the Carnot value. However, when the system operates as an engine, its efficiency is always smaller than the Carnot efficiency. As we will see in the next section, this situation will be reversed when the coherence is in the hot bath.

\subsection{Coherence in the hot bath}
We now consider the case where the coherence is in the hot bath.
We thus assume $T_1>T_2$. The conditions for the different operating regimes can be found by looking again at the signs of Eqs.~\eqref{eq:Q11}-\eqref{eq:W1} as we did in the previous section and are reported in Table~\ref{tab:functionings2}. Notice that the signs of $\dot Q_1$ and $\dot Q_2$ are reversed compared to Table~\ref{tab:functionings}.
\begin{table}[h]
\caption{Functionings for $T_1>T_2$.}
\begin{center}
\begin{tabular}{|c|c|}
\hline
$\dot W<0, \dot Q_1>0,\dot Q_2<0$ & engine
\\
\hline
$\dot W>0, \dot Q_1>0,\dot Q_2<0$ & accelerator
\\
\hline
$\dot W>0, \dot Q_1<0,\dot Q_2>0$ & refrigerator
\\
\hline
\end{tabular}
\end{center}
\label{tab:functionings2}
\end{table}

\begin{figure*}[t]
\begin{center}
\includegraphics[height=3.7cm]{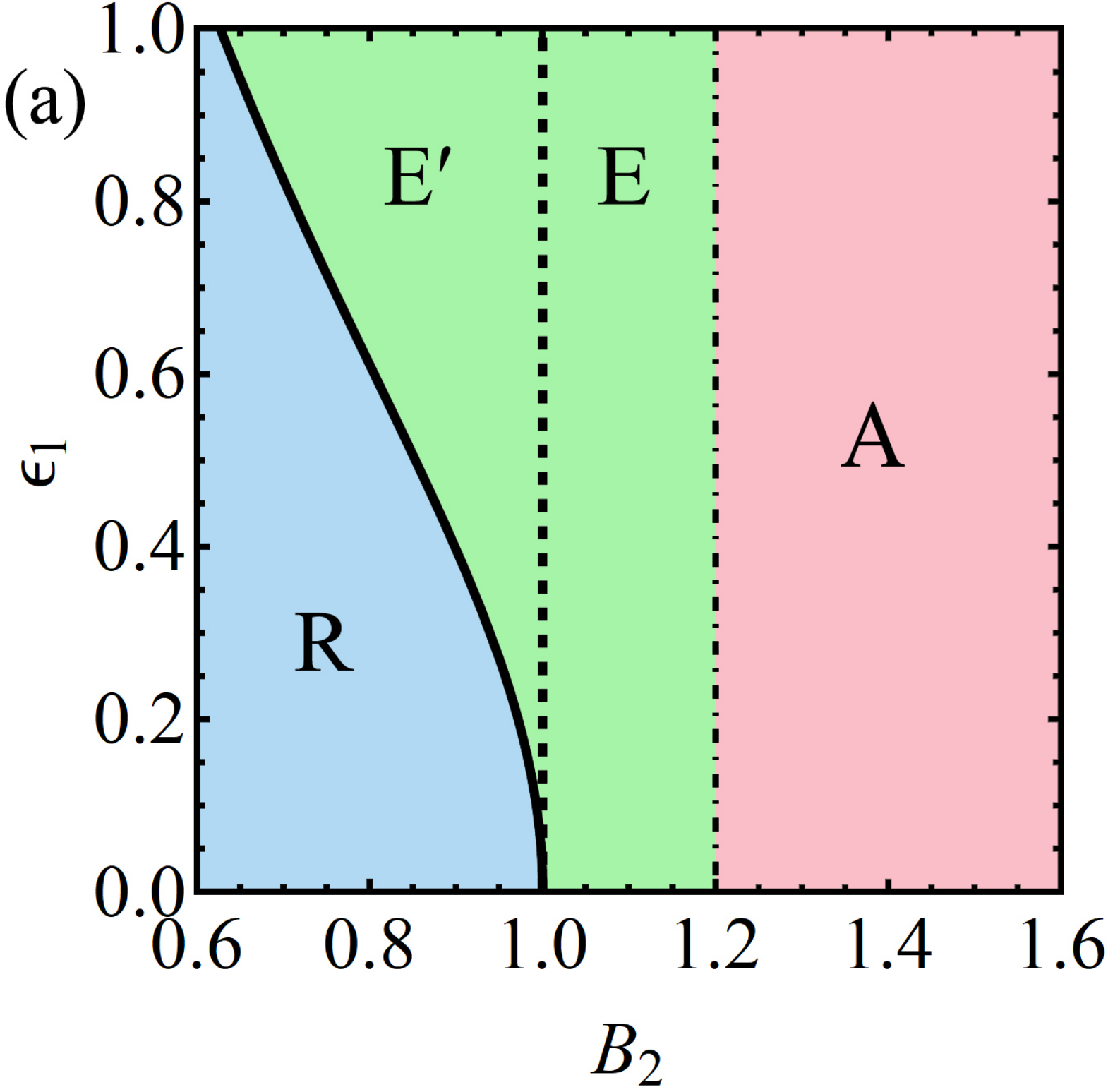}
\includegraphics[height=3.6cm]{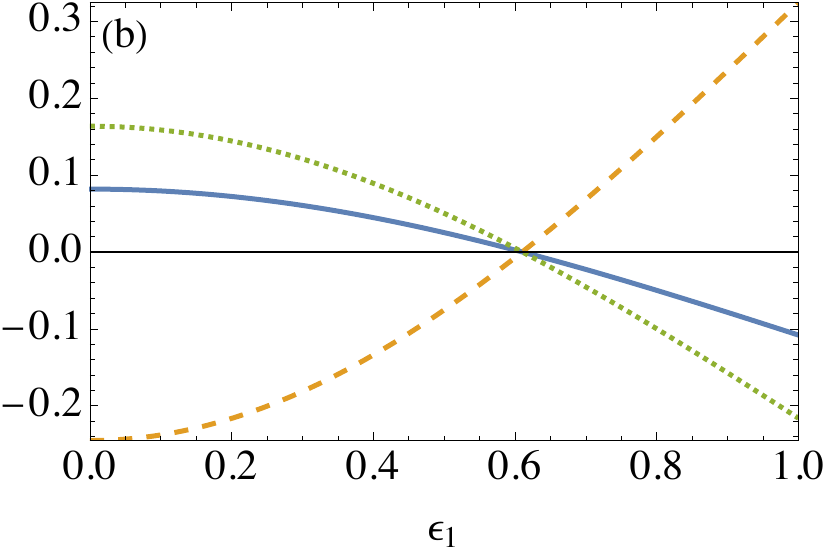}
\\
\includegraphics[height=3.6cm]{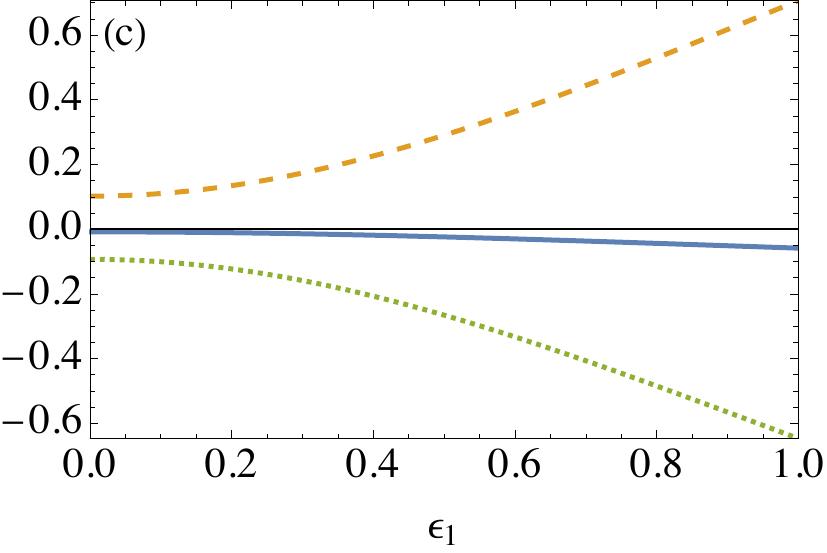}
\includegraphics[height=3.6cm]{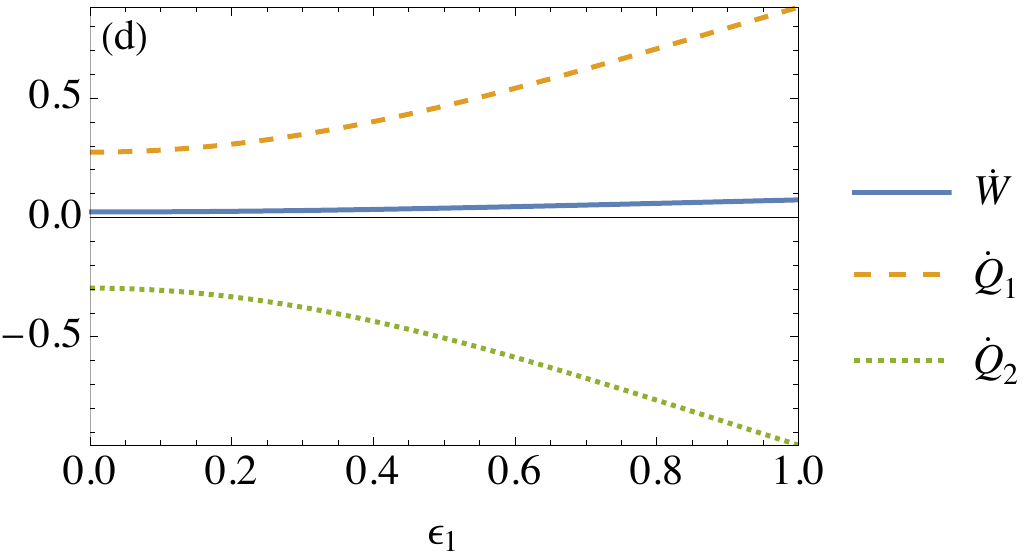}
\caption{Coherence in the hot bath: (a) Operational diagram for the different functionings of the thermal machine. Colour coding is as in Fig.~\ref{fig:coldbathcoherence}. In the area denoted as E', the device works as an engine with an efficiency larger than the Carnot value  The vertical solid line indicates the condition $n_1=n_2$. Heat currents and  power for (b) $B_2=0.8$, (c) $B_2=1.1$, (d) $B_2=1.3$. Other parameters: $B=1, B_1=1.2, T_1=3,T_2=2.5, \gamma=1$.}
\label{fig:hotbathcoherence}
\end{center}
\end{figure*}

\begin{figure}[t]
\begin{center}
\includegraphics[height=5cm]{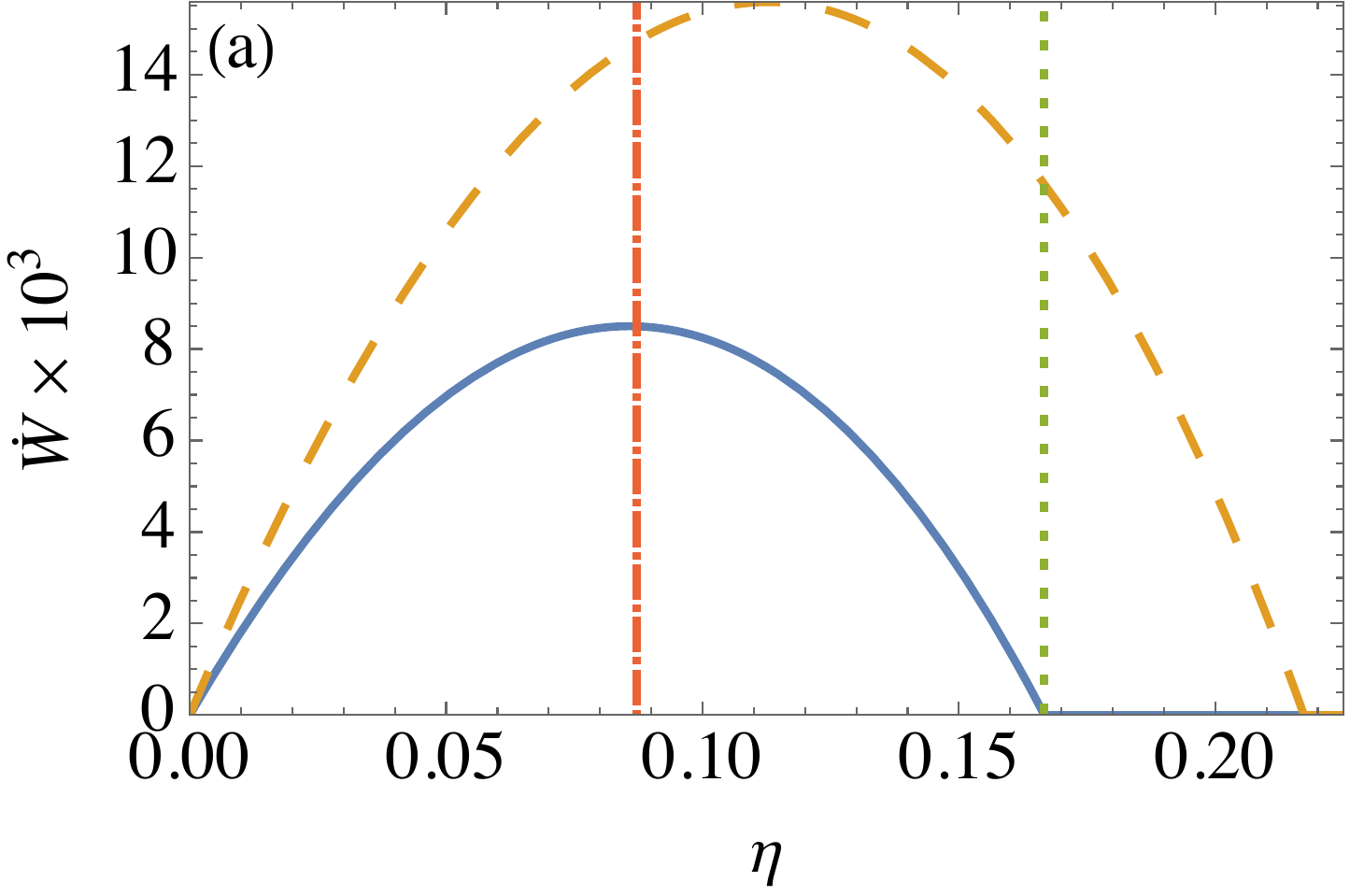}
\includegraphics[height=5cm]{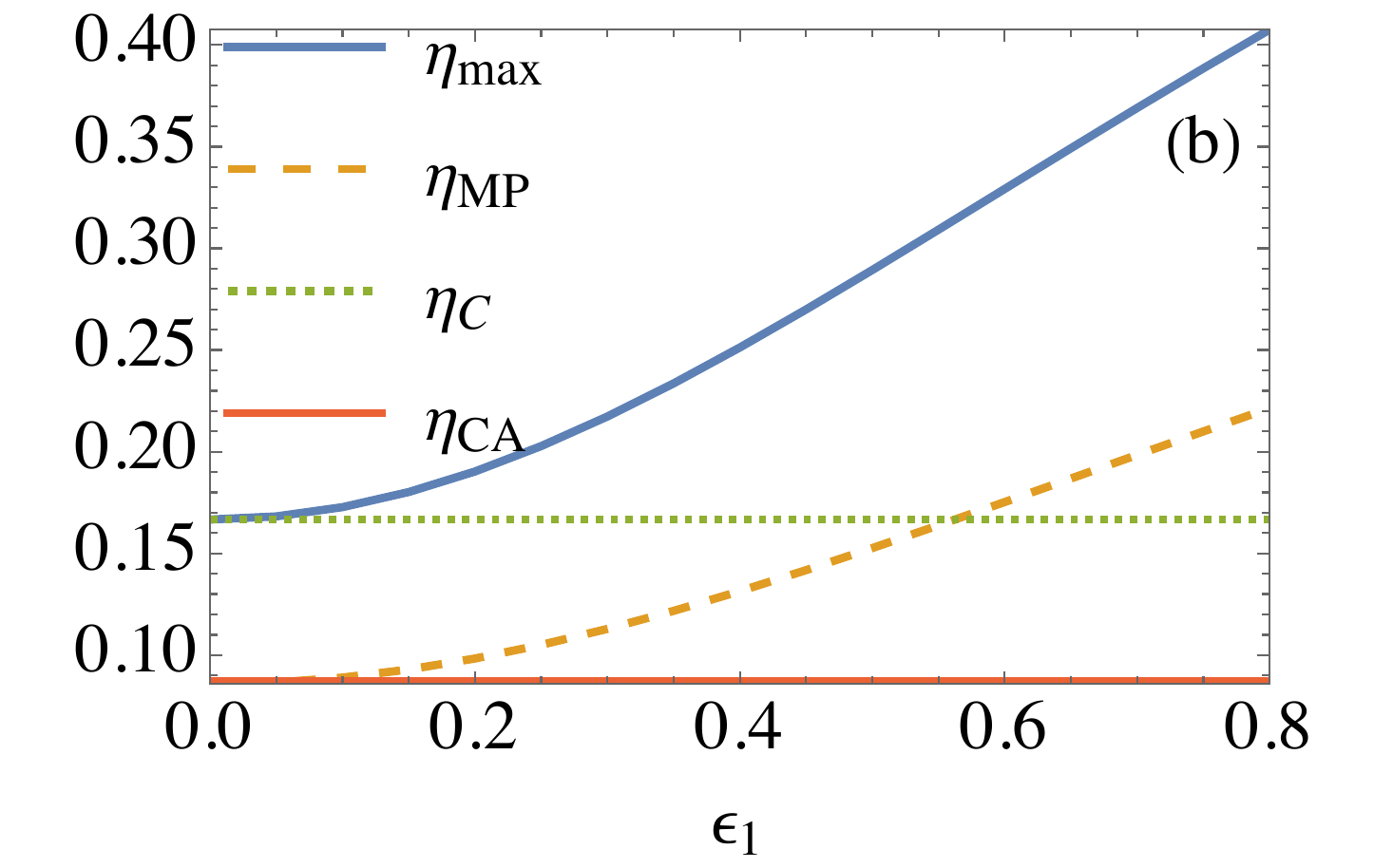}
\caption{(a) Power versus efficiency when the system is operating as an engine with the magnetic field $0.93<B_2<1.2$.  The solid (dashed) line is the case with $\epsilon_1=0$ ($\epsilon_1=0.1$). The vertical dotted line corresponds to the Carnot value $\eta_C$ (see Eq.~\eqref{eq:carnot}) and the dot-dashed line corresponds to the Curzon-Ahlborn value $\eta_{CA}$ (see Eq.~\eqref{effca}). (b): Maximum efficiency $\eta_{\rm max}$ (solid) and efficiency at maximum power $\eta_{MP}$ (dashed) as a function of the coherence in the hot bath. The two horizontal lines correspond to $\eta_{CA}$ (solid) and $\eta_C$ (dotted).  Other parameters as in Fig.~\ref{fig:hotbathcoherence}.}
\label{fig:effmaxpower}
\end{center}
\end{figure}

 Equations \eqref{eq:condition1}-\eqref{eq:condition2} are still valid and give us the conditions at which the power is zero. Using these equations we find the functioning diagram shown in Fig.~\ref{fig:hotbathcoherence}(a). In contrast to the case where the coherence is in the cold bath, we see that the device never operates as a hybrid refrigerator. Moreover, a coherence $\epsilon_1>\epsilon_1^*$ allows the system to operate as an engine even when $B_2<B_1$, where in absence of coherence a refrigerator would be expected. In this forbidden zone, the efficiency of the corresponding engine is larger than the Carnot value as shown in Fig.~\ref{fig:effmaxpower}.

In Fig.~\ref{fig:effmaxpower}(a), we show the power output $\dot W$ against the efficiency when the system behaves as an engine. In absence of coherence, $\epsilon_1=0$, the maximum achievable efficiency is the Carnot value $\eta_C$ (Eq.~\eqref{eq:etac}) where however the power is zero. The value of the efficiency at maximum power is obtained at the Curzon-Ahlborn value:
\begin{equation}\label{effca}
\eta_{CA} = 1-\sqrt{\frac{T_1}{T_2}}.
\end{equation}
On the other hand, in the presence of coherence  $\epsilon_1\neq 0$, the efficiency is much larger and surpasses both the Carnot value (at non zero power) and the Curzon-Ahlborn value. Fig.~\ref{fig:effmaxpower}(b), shows that the efficiency at maximum power $\eta_{MP}$ grows quadratically for very small $\epsilon_1$ and linearly for larger values. For a given $\epsilon_1$, the maximum value of the efficiency $\eta_{\rm max}$ is obtained for the value of $B_2$ at the transition, in the diagram of Fig.~\ref{fig:hotbathcoherence}(a), between the functioning as an engine and that as a refrigerator. The value of $B_2$ can be obtained by solving the equation:
\begin{equation}
\epsilon_1=\epsilon_1^*
\end{equation}
where $\epsilon_1^*$ is defined in Eq.~\eqref{eq:condition2}. The value of $\eta_{\rm max}$, plotted in Fig.~\ref{fig:effmaxpower}(b), is not universal, depending on all parameters, but represents an effective Carnot bound on the functioning of the machine as an engine.

\section{Summary and Conclusions}
\label{sec:conclusion}

In this paper, we have demonstrated the role of non-thermal bath effects in non-equilibrium processes. We provide a thermodynamic analysis of a microscopic collision model wherein a single qubit is interacting repeatedly with two local reservoirs at different temperatures, which consist of a collection of initially prepared ancillas with an infinitesimally small amount of coherence. 

A key insight is that the system-ancilla interaction does not satisfy local energy conservation, thus generating coherent and incoherent contributions in the heat flow and coherent and collisional terms in the power. In the continuous time limit, we have shown that the loss or gain of coherence in the state of the ancillas is given by a modified second law of thermodynamics which is described by a lower bound for the coherent heat current from one of the baths.

We have shown coherence to be a tuning parameter alongside the magnetic field of the baths for characterising the different operational regimes of the machine. Since including coherence in both baths can be reduced to the case with coherence in only one bath, to simplify the analysis, we studied the latter case.
We found that injecting some amount of coherence into the cold bath allows the refrigerator to survive the classically forbidden regime. This implies that its coefficient of performance surpasses the Carnot limit of the corresponding equilibrium reservoirs without coherence. 
In addition, coherence can result in advantageous effects such as the appearance of a hybrid refrigerator that simultaneously produces work and refrigerates the cold reservoir. In contrast, when the hot bath contains coherence and the cold reservoir is an equilibrium bath, the machine never operates as a hybrid refrigerator. In the case of the heat engine regime, coherence acts as fuel that drives the efficiency beyond the Carnot value corresponding to incoherent baths. Similarly, coherence leads to efficiency at maximum power much larger than the classical Curzon-Ahlborn value.

The simplicity of our model provides a general insight into the advantages of employing coherent reservoirs in the performance of thermodynamic tasks. Our findings can be further explored in systems of higher dimensionality, including quantum harmonic oscillators. Implementing our proposal in current experiments could face critical technical challenges such as scaling, decoherence and manufacturing errors. Nonetheless, the first proof-of-principle demonstrations have already shown great potential for future thermodynamic applications \cite{GarciaPerez2020,MeloPRA2022,Cattaneo2022}.

\ack
We thank Gonzalo Manzano, Marco Cattaneo and Gabriel Landi for interesting discussions and providing useful comments. We acknowledge the support by the UK EPSRC EP/S02994X/1 and UK EPSRC 2278075. K.H. and Y.H. would like also to thank the ICTP Office of External Activities (OEA) for the financial support through AF-14.

\appendix

\section{Derivation of the general expressions of heat and work }
\label{app:derivations}

The heat flow from bath $i$ is defined as the energy change in the bath's ancilla during a collision
\begin{equation}\label{eq:heat}
\dot{Q}_{i} = -\lim_{\tau \rightarrow 0} \frac{1}{\tau} \tr \Big[ H_{E,i} \Delta \rho \Big] ,  
\end{equation}
where
\begin{equation}
\Delta \rho = e^{-i\tau H_{\text{tot}}}\rho_S\otimes \rho_E e^{i\tau H_{\text{tot}}} - \rho_S\otimes \rho_E.
\end{equation}
We may write
\begin{eqnarray}
e^{-i\tau H_{\text{tot}}}\rho_S\otimes \rho_E e^{i\tau H_{\text{tot}}} = 
\rho_S\otimes \rho_E - i\tau \comm{H_{\text{tot}}}{\rho_S\otimes \rho_E} 
\\
- \frac{\tau^2}{2} \comm{H_{SE}}{\comm{H_{SE}}{\rho_S\otimes \rho_E}} + O(\tau^2)
\end{eqnarray}
therefore,
\begin{equation}\label{eq:deltarho}
\Delta \rho =   -i\tau \comm{H_{\text{tot}}}{\rho_S\otimes \rho_E} - \frac{\tau^2}{2} \comm{H_{SE}}{\comm{H_{SE}}{\rho_S\otimes \rho_E}} + O(\tau^2).
\end{equation}
In the following we will ignore terms of order $\tau^2$ and higher, as they will tend to zero when dividing by $\tau$ and taking the limit as $\tau \rightarrow 0$. Plugging Eq. \eqref{eq:deltarho} into \eqref{eq:heat}, we find
\begin{equation}\label{eq:heatparts}
\dot{Q}_i = i \lim_{\tau \rightarrow 0} \tr\Big[ H_{E,i} \comm{H_{\text{tot}}}{\rho_S\otimes \rho_E} \Big] + \frac{1}{2} \lim_{\tau \rightarrow 0}  \tr\Big[ \tau  H_{E,i} \comm{H_{SE}}{\comm{H_{SE}}{\rho_S\otimes \rho_E}} \Big].
\end{equation}
The first term represents the coherent heat flow, defined in Eq.~\eqref{eq:Qdotcoherent}:
\begin{eqnarray}
 \dot{Q}_{i,\text{coh}}&=&i \lim_{\tau \rightarrow 0} \tr\Big[ H_{E,i} \comm{H_{\text{tot}}}{\rho_S\otimes \rho_E} \Big]  \\
&=& i \lim_{\tau \rightarrow 0} \tr\Big[ \comm{H_{E,i}}{H_{SE_i}} \rho_S\otimes\rho_E \Big]\\
&=& - i \lim_{\tau \rightarrow 0} \tr\Big[ \comm{G_{E,i}}{H_{E,i}} \rho_{E,i} \Big],
\end{eqnarray}
where $G_{E,i} $ is defined in Eq.~\eqref{eq:GEi}.

The second expression in Eq. \eqref{eq:heatparts} therefore gives rise to the  incoherent heat flow
\begin{eqnarray}
\dot{Q}_{i,\text{inc}}& =& \frac{1}{2} \lim_{\tau \rightarrow 0}  \tr\Big[ \tau  H_{E,i} \comm{H_{SE}}{\comm{H_{SE}}{\rho_S\otimes \rho_E}} \Big]\\
&=& \frac{1}{2} \lim_{\tau \rightarrow 0}  \tr\Big[\tau \comm{H_{SE}}{\comm{H_{SE}}{H_{E,i}}} \rho_S\otimes\rho_E\Big].
\end{eqnarray}

Now let us pass to work. The power is defined as
\begin{equation}\label{eq:power}
\dot{W} = \left\langle \frac{\partial H_{\text{tot}}}{\partial t}\right\rangle=    \lim_{\tau \rightarrow 0} \frac{1}{\tau} \tr\Big[ (H_S +H_E) \Delta \rho \Big].
\end{equation}
Plugging in Eq. \eqref{eq:deltarho}, we find
\begin{eqnarray}
\label{eq:powerparts1}
\dot{W}&=& -i \lim_{\tau \rightarrow 0} \tr \Big[ (H_S +H_E) \comm{H_{\text{tot}}}{\rho_S\otimes \rho_E} \Big] \\
\label{eq:powerparts2}
&-& \frac{1}{2} \lim_{\tau \rightarrow 0}  \tr\Big[ \tau  (H_S +H_E) \comm{H_{SE}}{\comm{H_{SE}}{\rho_S\otimes \rho_E}} \Big].
\end{eqnarray}
The expression in Eq. \eqref{eq:powerparts1} gives rise to coherent power
\begin{eqnarray}
 \dot{W}_{\text{coh}}&=& -i \lim_{\tau \rightarrow 0} \tr\Big[ (H_S +H_E) \comm{H_{SE}}{\rho_S\otimes \rho_E} \Big]  \\
&=& i \lim_{\tau \rightarrow 0} \tr\Big[ \comm{ H_{SE}}{H_S + H_E} \rho_S\otimes\rho_E \Big]
\end{eqnarray}
The collisional power then arises from the second term in Eq. \eqref{eq:powerparts2}
\begin{eqnarray}
 \dot{W}_{\text{col}}&=& -\frac{1}{2} \lim_{\tau \rightarrow 0}  \tr\Big[ \tau  (H_S + H_E) \comm{H_{SE}}{\comm{H_{SE}}{\rho_S\otimes \rho_E}} \Big]\\
&=& -\frac{1}{2} \lim_{\tau \rightarrow 0}  \tr\Big[\tau \comm{H_{SE}}{\comm{H_{SE}}{H_S + H_E}} \rho_S\otimes\rho_E\Big].
\end{eqnarray}

\section{Steady state solution}
\label{sec:appendix}

In this section, we show the explicit expression of the system's steady state when the system is subject to the master equation Eq.~\eqref{eq:mequbit}. We write the density matrix $\rho_S$ in the basis of eigenstates of $\sigma^z_S$ as:
\begin{equation}
\rho_S = \left(\begin{array}{cc}\rho_{11}& \rho_{12} \\ \rho_{21} & \rho_{22}\end{array}\right).
\end{equation}
In this representation, the entries of the steady state density matrix read:
\begin{eqnarray}
\label{eq:ss1}
\rho_{11}&=&\frac{1}{R}\left\{ 4B^2 n+\gamma_{\rm eff}(1+2n)^2(2\epsilon_{\rm eff}^2+n \gamma_{\rm eff})\right\},
\\
\label{eq:ss2}
\rho_{22}&=&1-\rho_{11},
\\
\label{eq:ss3}
\rho_{21} &=&\frac{ \ii \epsilon_{\rm eff} e^{i\phi}\sqrt{2\gamma_{\rm eff}(2n+1)}}{R}\left\{ 
2i B + (2n+1)\gamma_{\rm eff} \right\},
\nonumber \\
\\
R&=& (2n+1)\left [4B^2 + \gamma_{\rm eff} (2n+1)(4\epsilon_{\rm eff}^2 +(2n+1)\gamma_{\rm eff} )    \right], \nonumber\\
\end{eqnarray}
where we have introduced the effective decay rate $\gamma_{\rm eff} =2\gamma$ and  an effective coherence strength, expressed in complex polar form:
\begin{equation}
\epsilon_{\rm eff} e^{i\phi} =\frac{ \epsilon_1 e^{i \phi_1}
   \sqrt{1 +2   n_1}+\epsilon_2 e^{i \phi_2} \sqrt{1 +2  n_2} }{\sqrt 2\sqrt{1+2n}},
\end{equation}
with $\epsilon_{\rm eff}$ and $\phi$, real parameters, denoting its magnitude and phase, respectively.

\section{Thermodynamic quantities: detailed expressions}
\label{sec:appendixB}
Here we report the expressions for the coherent and incoherent heat currents as well as those for the power. 

Splitting the heat flow into its coherent and incoherent parts as in Eqs.~\eqref{eq:Qdotcoherent} and \eqref{eq:Qdotincoherent} we find 
\begin{eqnarray}
\label{eq:Qdotcoherentappendix}
\dot{Q}_{1,\text{coh}} &=& \ii \epsilon_1  B_1  \sqrt{2\gamma (1 + 2n_1)} (e^{i \phi} \rho_{12} - e^{-i \phi} \rho_{21}),
\\
\dot{Q}_{1,\text{inc}} &=& - 4 B_1 \gamma \left[\rho_{11} +n_1 (\rho_{11}-\rho_{22})\right],
\end{eqnarray}
for the first ancilla, initially coherent, and 
\begin{eqnarray}
\dot{Q}_{2,\text{coh}} &=& 0, \nonumber\\
\dot{Q}_{2,\text{inc}} &=&- 4 B_2 \gamma \left[\rho_{11} +n_2 (\rho_{11}-\rho_{22})\right].
\end{eqnarray}
for the second ancilla, initially in an incoherent thermal state.

We find the coherent and collisional parts of the power to be
\begin{eqnarray}
\dot{W}_{\text{coh}} &=&  2 \epsilon_1 \ii (B-B_1)  \sqrt{2\gamma (1 + 2n_1)} (e^{i \phi} \rho_{12} - e^{-i \phi} \rho_{21}),
\\
\dot{W}_{\text{col}} &=& -4 \gamma \left\{ [(B-B_1)(1+n_1)+(B-B_2)(1+n_2)]\rho_{11} \right .
\\
&-& \left . [(B-B_1)n_1+(B-B_2)n_2]\rho_{22} \right\}. \nonumber
\end{eqnarray}

Steady state expressions can be directly obtained by replacing the stationary solutions Eqs.~\eqref{eq:ss1}-\eqref{eq:ss3} in the relevant expressions for heat and power.

The rate of change of the relative entropy of coherence in the first ancilla is
\begin{eqnarray}
\fl \dot{\mathcal{C}}(\rho_{E,1})  = -\beta_1 B_1 \epsilon_1^2 \gamma^2 (1+2n_1) 
\left\{\left[ 1+2n_1^2+4n_1(1+n_2)+2n_2(2+n_2) \right] \times \right.
\nonumber \\
\times\left[ B^2 +\gamma^2 (1+n_1+n_2)^2 \right]
+\left . 2 \epsilon^2 \gamma (1+2n_1)(1+n_1+n_2)^2
\right\}
\\
/
\left\{ (1+n_1+n_2)^2 \left[B^2 +\gamma^2 (1+n_1+n_2)^2 + \epsilon^2 \gamma(1+2n_1) \right]^2
\right\}, \nonumber
\end{eqnarray}
which is always negative
and in the second ancilla is
\begin{equation}
\dot{\mathcal{C}}(\rho_{E,2})  =  
\frac{\beta_2 B_2\epsilon_1^2 \gamma^2 (1+2n_1) [B^2 +\gamma^2 (1+n_1+n_2)^2] }{(1+n_1+n_2)^2 \left[ B^2 +\gamma^2 (1+n_1+n_2)^2 + \epsilon^2 \gamma(1+2n_1) \right]^2}.
\end{equation}
which is always positive.

\section{Quantum coherence in both baths}
\label{sec:appendixC}

In this appendix we analyse what happens in the case in which coherence is present in both baths. We are going to show that the expressions of the heat currents and power can be also obtained with coherence only in one of the two baths and with an effective amplitude.

We consider the master equation Eq.~\eqref{eq:mequbit} for both $\epsilon_1\neq 0$ and $\epsilon_2 \neq 0$. After finding the corresponding steady state we obtain the following energy flows 
\begin{equation}
\dot{Q}_1 = B_1 V^{(2)}(\epsilon_1,\phi_1,\epsilon_2,\phi_2),
\end{equation}
\begin{equation}
\dot{Q}_2 = -B_2 V^{(2)}(\epsilon_1,\phi_1,\epsilon_2,\phi_2),  
\end{equation}
\begin{equation}
\dot{W} = (B_2 - B_1) V^{(2)}(\epsilon_1,\phi_1,\epsilon_2,\phi_2),   
\end{equation}
which are similar to those in Eqs.~\eqref{eq:Q11}-\eqref{eq:W1}
but now the common factor depends on all the parameters, $\epsilon_1$,$\phi_1$, $\epsilon_2$, and $\phi_2$:
\begin{eqnarray}
\label{eq:V12}
\fl V^{(2)}(\epsilon_1,\phi_1,\epsilon_2,\phi_2)&=&
2\gamma\left\{B^2(n_1-n_2)+ \gamma(1+n_1+n_2) \right.
\nonumber\\
&\times&\Big[\gamma n_1 (1 + n_1) - 
\gamma n_2 (1 + n_2) + \epsilon_1^2(1 + 2 n_1)  - \epsilon_2^2(1 + 2 n_2)\Big]  
\nonumber\\
&+& \left .2B \epsilon_1 \epsilon_2\sqrt{(1+2n_1)(1+2n_2)}  \sin(\phi_1 - \phi_2) \right \}
\nonumber \\
&/&\left\{(1 + n_1 + n_2) \left[B^2 + \gamma^2 (1 + n_1 + n_2)^2 \right .
+\gamma \epsilon_1^2(1+2n_1) +\gamma \epsilon_2^2(1+2n_2)\right] 
\nonumber \\
&+&\left . 2\gamma\epsilon_1 \epsilon_2\sqrt{(1 + 2 n_1) (1 + 2 n_2)} \cos(\phi_1 - \phi_2)\right\}.
\end{eqnarray}

Let us now consider the corresponding expressions for heat currents and power when coherence is only in one bath, Eqs.~\eqref{eq:Q11}-\eqref{eq:W1}. These are all proportional to a common factor $V(\epsilon_1)$, reported in Eq.~\eqref{eq:V1}, that only depends on the coherence amplitude $\epsilon_1$ and not its phase $\phi_1$. If we now choose $\epsilon_1=A_1/A_2$ where
\begin{eqnarray}
\fl A_1^2 = \Big(B^2 + \gamma^2(1 + n_1 + n_2)^2\Big) \Bigg[\gamma(1 + 2 n_1) (1 + 2 n_2)  (\epsilon_1 - \epsilon_2) (\epsilon_1 +\epsilon_2)
\\
+ 2\epsilon_1 \epsilon_2 \sqrt{(1 + 2 n_1) (1 + 2 n_2)}  \Big(\gamma(n_2 - n_1)\cos(\phi_1 - \phi_2) + B \sin(\phi_1 - \phi_2)\Big)\Bigg] ,
\nonumber
\\
\fl A_2^2 = \gamma(1 + 2 n_1)\Bigg[(1 + 2 n_2) \Big(B^2 + \gamma^2(1 + n_1 + n_2)^2 + 2\gamma (1 + n_1 + n_2)\epsilon_2^2)\Big)
\\
+ 2\epsilon_1 \epsilon_2 \sqrt{(1 + 2 n_1) (1 + 2 n_2)}  \Big(\gamma(1 + n_1 + n_2)  \cos(\phi_1 - \phi_2) - B \sin(\phi_1 - \phi_2)\Big)\Bigg],
\nonumber
\end{eqnarray}
then the common factor for the case with coherence in the two baths is equal to the common factor for the case with coherence in only one bath:
\begin{equation}
V(A_1/A_2) = V^{(2)}(\epsilon_1,\phi_1,\epsilon_2,\phi_2).
\end{equation}
As a consequence all thermodynamic quantities that arise in the case with coherence in both baths can be also reproduced with coherence only in a single bath.


\section*{References}
\bibliographystyle{iopart-num}
\bibliography{biblio2}
\end{document}